# Nonlocal supercurrent of quartets in a three-terminal Josephson junction

**Authors:**

*Yonatan Cohen[†,1], Yuval Ronen[†,1], Jung-Hyun Kang[1], Moty Heiblum[#,1], Denis Feinberg[2,3], Régis Mélin[2,3], and Hadas Shtrikman[1]*

**Affiliation:**

[1] *Braun Center for Submicron Research, Department of Condensed Matter Physics, Weizmann Institute of Science, Rehovot 76100, Israel*

[2] *CNRS, Institut NEEL, F-38042 Grenoble, France*

[3] *Université Grenoble-Alpes, Institut NEEL, F-38042 Grenoble, France*

[†] *Equal contributions*

[#] *Corresponding Author (moty.heiblum@weizmann.ac.il)*

## Abstract

We report an observation of a new, non-dissipative and non-local supercurrent, carried by *quartets;* each consisting of four entangled electrons. The supercurrent is a result of a novel Andreev bound state (ABS), formed among three superconducting terminals. While in a two-terminal Josephson junction the usual ABS, and thus the DC Josephson current, exist only in equilibrium, in the present realization the ABS exists also in the strongly nonlinear regime (biased terminals). The presence of supercurrent carried by quartets was established by performing non-local conductance and cross-correlation of current fluctuations measurements, in different devices made of aluminum-InAs nanowire junctions. An extensive and detailed theoretical study is intertwined with the experimental results.

# Introduction

Superconductivity is one of modern physics' triumphs, manifesting a macroscopic phenomenon governed by quantum mechanics, stressing the significance of the 'phase' of a macroscopic wave function [1]. Most striking is the 'DC Josephson effect' [2]: In response to a phase-difference between two superconductors (SCs) connected via a 'weak link', a non-dissipative supercurrent flows through the junction in equilibrium (Fig. 1(a)). Moreover, biasing the junction drives an evolution of the phase difference with time, leading to an oscillatory supercurrent: the 'AC Josephson effect'. Here, we demonstrate the formation of the non-equilibrium DC Josephson current (Fig. 1(b)), being a consequence of paired Cooper pairs, so-called *quartets* [3].

In unbiased three-terminal Josephson junction two-terminal supercurrent of Cooper pairs flows from any one terminal to another. Away from equilibrium, these DC supercurrents vanish. Yet under certain biasing conditions, new type of supercurrents may emerge. The simplest one emerges when $V_L=-V_R$, with both voltages applied with respect to the third, usually grounded, terminal, $S_M$ (Fig. 1(b)). Under this condition, it is predicted that two Cooper pairs, one emerging from terminal $S_R$ and the other from terminal $S_L$, interact in the junction to form a Quartet in $S_M$; namely, a quasiparticle composed of four entangled electrons [3-6]. As shown in Fig. 1(b), this can happen only if terminal $S_M$ is narrow (size $L$) in comparison to the superconducting coherence length $\xi$, thus allowing formation of Cooper pairs via 'crossed Andreev reflection' (CAR). Evidently, the reversed process should also take place; where two Cooper pairs in $S_M$ split (each of them via CAR, [7]) and form two, spatially separated, entangled Cooper pairs in terminals $S_L$ and $S_R$. As will be discussed below, this new supercurrent is *nonlocal* in the sense that the current from, say, $S_M$ to $S_R$ depends on the phase of $S_L$.

A previous study of the conductance in a three-terminal metallic junction provided a signature of the formation of quartets [6]. However, several alternative models prevented a clear conclusion of the origin of this effect. Here, we provide conclusive evidence of the nonlocal nature of the supercurrent, and its formation by quartets. The presented measurement results contradict alternative explanations for the observed supercurrent. Note, also, that we observed higher order non-dissipative supercurrents, such as *sextets* [4,5] and new dissipative quasiparticle channels, such as *nonlocal* multiple Andreev reflections (MAR) [8-10].

**The quartet supercurrent**

The microscopic picture of supercurrent flow in a short two-terminal superconductor-normal-superconductor (SNS) Josephson junction (JJ) is shown in Fig. 1(c). An electron impinging at the superconducting gap is reflected back as a hole via Andreev reflection (AR); hence, transmitting a Cooper pair into the superconductor. An Andreev bound state (ABS), formed between the two superconductors carries the equilibrium supercurrent. The magnitude of the supercurrent obeys the energy-phase relation, $I = -\frac{2e}{\hbar}\frac{dE_{ABS}}{d(\Delta\varphi)}$, with $E_{ABS}$ the energy of the ABS and $\Delta\varphi$ the phase difference between the two superconductors [11].

The microscopic picture of the quartet supercurrent flow in a three-terminal JJ is shown in Fig. 1(d) [12-14]. Due to CAR processes in the narrow terminal $S_M$, an outgoing hole, in response to an incoming electron, propagates towards the opposite terminal, and a new ABS that connects all three superconducting terminals is formed [3]. The ABS's energy, $E_{ABS}(\varphi_L, \varphi_R)$, is a function of two independent phases, $\varphi_L$ and $\varphi_R$, each with respect to the phase of the center terminal with $\varphi_M = 0$. Since, the ABS of the quartet exists under asymmetric biasing conditions, $V=V_R=-V_L$ (Fig. 1(d)), it is beneficial to choose new variables, $E_{ABS}(\varphi_q, \chi)$, where $\varphi_q = \varphi_L + \varphi_R$ and $\chi = \varphi_L - \varphi_R$, with the phase $\varphi_q$ is stationary while the phase $\chi$ is continuously changing, $\chi = \frac{4e}{\hbar}Vt$. In a semi-classical picture, one may average the energy over time, $\langle E_{ABS}(\varphi_q, \chi)\rangle_t$, yielding a $\varphi_q$ dependent effective energy $E_{ABS,eff}(\varphi_q)$, with a supercurrent $I_{quartet} = -\frac{2e}{\hbar}\frac{dE_{ABS,eff}}{d\varphi_q}$ (Supplementary Information, S1).

Considering the geometry shown in Fig. 1(b), the probability amplitude for a CAR process is expected to be large if $L < \xi \approx 0.2 - 0.3 \mu m$. Indeed, previous experiments have clearly shown the presence of a dominant CAR process in similar configurations based on similar InAs nanowire devices [7,20].

## Quantum noise

'Particle-hole symmetry' dictates that ABS's should appear in pairs of particle–hole conjugates. The applied voltage *V*, makes the phase difference between the terminals grow linearly with time, and thus the energy of the ABSs oscillates. As shown in Fig. 1(e), Landau-Zener (LZ) transitions between the two quartet ABSs may take place as the energy gap between them quenches [15,16]. Void of such transitions the Josephson current is expected to be noiseless. However, transitions between the states introduce stochasticity in the occupation of the ABSs [17, 18], leading to strong current fluctuations (see Supplementary Information, S1 (a)&(b)). It is predicted that the resulting cross-correlation of current fluctuations between $S_L$ and $S_R$, would be positive, since the quartet ABS [24] carry supercurrent which flows simultaneously into (or out of) these two terminals. This could also eliminate some of the alternative explanations, as discussed below. Modeling the non-equilibrium dynamics in a similar three-terminal structure (Supplementary Information, S1), shows the dependence of the cross-correlation (CC) as a function of *V*, reflecting the nature of LZ transitions. The CC depends on the stationary quartet phase in a non-monotonic characteristic fashion which is also observed experimentally.

## Experimental setup

Three different configurations of the three-terminal JJ were realized by coupling aluminum superconducting contacts to InAs nanowires: *Device d1* - a single nanowire configuration with three terminals along the nanowire with the central contact, $S_M$, being narrower than the superconductor coherence length (Fig. 2(a)). This configuration suppresses direct coupling between $S_L$ and $S_R$, and increasing CAR via $S_M$. *Device d2* - a single nanowire configuration with a wide central contact, $S_M$ (Fig. 2(b)). This configuration suppresses CAR in $S_M$. *Device d3* - a Y-shape merged nanowires configuration, where each terminal communicates with the other two (Fig. 4(a)).

The nanowires were grown by a gold assisted MBE process, using the well-established vapor-liquid-solid growth technique. Growth was initiated on an un-patterned (100) InAs substrate, where both, single wires and the Y-shape intersections were formed [25]. Devices were fabricated on an oxidized P[+]-doped Si wafer (with 150nm thick $SiO_2$), with superconducting contacts and local gates made by depositing 5nm/120nm Ti/Al. The measurement setup allowed measuring the differential conductance and 'zero frequency' cross-correlation (CC) of current fluctuations in $S_L$ and $S_R$ (Supplementary Information, S2). We define $G_{L/R} = dI_{L/R}/dV_M$, where $I_{L/R}$ is the current in $S_L$ or $S_R$, and $V_M$ is a small AC signal applied to the central contact. The DC bias to $S_L$ and $S_R$, for the CC measurements, was applied on a 5Ω resistor at the source (Supplementary Information, S2). The induced superconducting energy gap in the nanowire was $2\Delta \approx 140 \mu eV$.

## Results and Discussion

### *Differential conductance measurements*

Figure 2(c) presents a color representation of $G_L$ as function of the applied biases $V_L$ and $V_R$ in *device d1* (equivalent plot of $G_R$ is shown in Fig. S2 (b)). A pronounced high conductance peak is observed for $V_L=-V_R$; agreeing with the expected signature of the *quartet*. Traces at $V_R=-16\mu V$ show $G_L$ and $G_R$ as function of $V_L$ in Fig. 2(e). The sharp peaks at $V_L=+16\mu V$ emphasize the difference between the quartet conductance peak and the broader peaks attributed to dissipative MAR processes. Moreover, the quartet's conductance peak is accompanied by two dips at its sides (inset, Fig. 2(e)); a typical fingerprint is also the ubiquitous 'zero-bias-conductance-peak' of the two-terminal JJ (see Fig. 2(e), and Supplementary Information, S1(c)) [24]. Calculating the phase dynamics around the quartet peak, with an 'effective quartet's RSJ model' (Supplementary Information, S1(b)), allows accessing a typical 'quartet energy' extracted from the peak's width $E_q$. Since $E_q \approx \frac{\hbar I_C}{2e} \approx 2\mu eV$, the critical quartet's supercurrent is ~0.6nA. Other, non-dissipative processes, which lead to conductance lines with different slopes, are visible. For example, a sextet line at $V_R=-2V_L$ (and $V_L=-2V_R$), represents a six-electron entangled state, which involves three Cooper pairs [4,5].

Similar measurements were performed on *device d2* − where $S_M$ is much longer than the coherence length of the aluminum superconductor. No sign of quartet, or any higher order supercurrent, was observed (Fig. 2(d)). This is a crucial test since the quartet interpretation for the line observed in *device d1* requires presence of CAR at electrode $S_M$.

## Cross-correlation of current fluctuations

The CC between current fluctuations in $S_L$ and $S_R$ was measured with $V_R$~15µV. A clear positive CC peak, coinciding with the quartet's conductance peaks, is observed (Fig. 3(a), lower panel). The small negative fluctuating background in the CC signal results from various MAR processes [20, 26-29]. The evolution of the CC signal strength along the quartet's conductance line (Figs. 3(b) & 3(c) upper panel), agrees qualitatively with a calculation which attributes the LZ transitions as the cause of the current fluctuations (SI, S1(b)). A quantitative estimate of the CC signal and its evolution with applied voltage is quite difficult to perform, since it depends on the detailed experimental conditions (Fig. 3(c) lower panel). It should be stressed that the positive CC signal excludes a MAR related signal, since the latter is expected to give a negative CC signal at the quartet biasing condition (S1(d)).

## Nonlocality of the quartet

Evidently, the formation of the quartet quasiparticle requires Cooper pairs arriving at $S_M$ from $S_L$ and $S_R$. Hence, suppressing $G_R$ should also suppress $G_L$ along the quartet's line. Indeed, pinching the right arm (with negative $V_{GR}$), quenches the quartet's line in both sides (Fig. 3(d)), as well as the respective CC signal (Fig. 3(d)).

## Are there alternative mechanisms for the $V_L$=-$V_R$ line?

Under the biasing condition, $V=V_L=-V_R$, the nanowire system is expected to generate two oscillating Josephson currents, with matching frequencies, $\dot{\varphi}_L = -\dot{\varphi}_R = \frac{2eV}{\hbar}$. Radiation emitted by one JJ (say R) can be absorbed by the other (L) - enabled by the electromagnetic environment. The electromagnetic environment of *device d2* is identical to that of *d1*. Since the photons have an energy of ~10µeV (wavelength ~1cm) we would expect the effect not to diminish, however in *device d2* the conductance line at $V_L$=-$V_R$ is missing. These results on *device d2* rule out this mechanism.

Another mechanism that may couple the two AC Josephson currents is a common resistive element in $S_M$ [30]. This scenario is in fact not relevant to the devices shown above as there is no such common element, comparable to the bare nanowire resistance. This is further explained and confirmed in the Supplementary Information, section S5.

*The role of the dissipative MAR processes*

It is worth addressing the effect of the non-local MAR process in a NW based three-terminal JJ as it is seen here for the first time as well as its effect on the quartet's fingerprints. The MAR processes can be *local*, with only two terminals taking part in the transport mechanism, or *nonlocal*, incorporating all three terminals [9]. They can be divided into two categories: (*L*1) $mV_L+nV_R=2\Delta$ and (*L*2) $mV_L+nV_R=0$, where $m$ and $n$ are integers (see Supplementary Information, S6). We present the measurement results with *device d3* (Y-shaped, Fig. 4(a)), where *local* and *nonlocal* MAR are observed. Since the MAR features are relatively faint, we plot the derivative of $G_L$ with respect to $V_L$ as a function of $V_L$ and $V_R$ (*i.e.* the second derivative of the current). While lines belonging to the *L*2 family were not observed, a rich sub-gap structure with certain lines belonging to the *L*1 family was observed (see guide lines in Fig. 4(b)). For example, lines that correspond to *nonlocal* MAR processes such as the (*m,n*=-2,1) and (*m,n*=-3,2), involving CAR processes, are highlighted in Figs. 4(c) & 4(d).

Comparing the conductance line of the quartet and the non-local MAR; as eluded above, the shape and width of the quartet's conductance peak should resemble that of the equilibrium Josephson peak and not the smooth and wide conductance peaks of MAR processes (see Supplementary Information, S6). Moreover, a *nonlocal* MAR process, in the relevant range of interest, is a tenth order process with small energy separation between adjacent peaks. It is likely also to experience inelastic scattering events and thus will inherently appear as a wide peak, sometimes overlapping with others. Note, that the only faintly observed non-local MAR peaks in *device d3* are of the lowest order. Finally, as mentioned earlier, the CC signal on the quartet line is positive, consistent with rapid Landau-Zener transitions between the quartet's ABS's, while the signature for CC associated with *nonlocal* MAR process (at $V_L=-V_R$) is expected to be negative.

**Summary**

We presented detailed study of a non-local, coherent, strongly non-equilibrium phenomenon, which results in supercurrent in a three-terminal Josephson junction. Necessitating CAR processes, a novel many-body quantum state is formed, with *quartets* quasiparticles − each made of two entangled Cooper pairs − carrying the supercurrent. Measurements of nonlocal conductance and cross-correlation of current fluctuations, performed on three types of three-terminal-devices, show a definite signature of quartets' supercurrent. Alternative mechanisms that may have produced similar effects were tested and disproved. We provide theoretical estimates that agree qualitatively with the measured quantities.

**Acknowledgement:**

D. F. and R. M. acknowledge support from ANR Nanoquartets 12-BS-10-007-04 and the CRIANN computing centre. M.H. acknowledges the partial support of the Israeli Science Foundation (ISF), the Minerva foundation, the U.S.-Israel Bi-National Science Foundation (BSF), the European Research Council under the European Community's Seventh Framework Program (FP7/2007-2013)/ERC Grant agreement No. 339070, and the German Israeli Project Cooperation (DIP). H.S. acknowledges partial support by ISF grant number 532/12, and IMOST grants #0321-4801 & #3-8668. H.S., incumbent of the Henry and Gertrude F. Rothschild Research Fellow Chair.

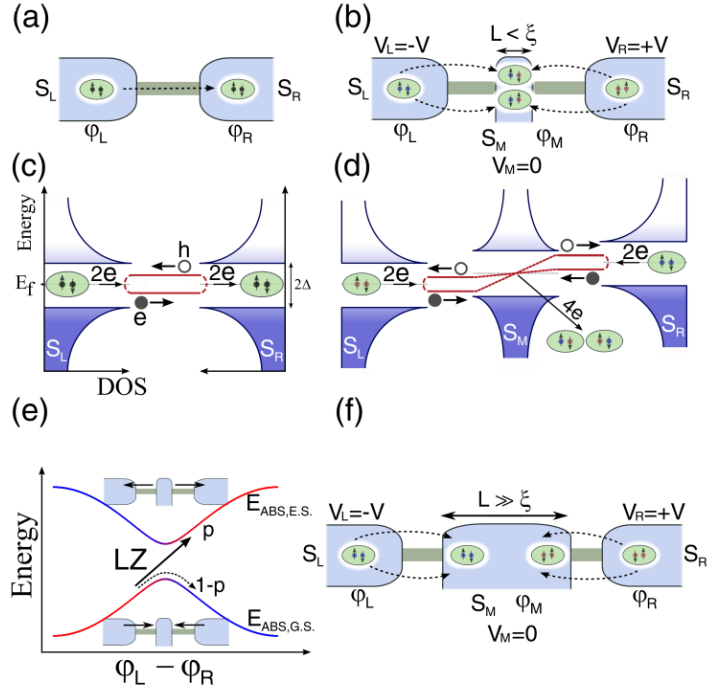

# Fig. 1

**Fig. 1. Non-dissipative current at two and three terminal Josephson junctions.**

(a) Schematic illustration of a two-terminal Josephson junction. (b) Schematic illustration of a three-terminal Josephson junction with a narrow central contact, and the formation of a quartet by entangling two distinct Cooper pairs. (c) Schematic illustration of the two-terminal resonance process of an Andreev Bound State (ABS), enabling Josephson supercurrent flow. (d) Schematic illustration of the three-terminal quartet ABS, leading to a nonlocal supercurrent flow. (e) Dependence of the two quartet particle-hole conjugates ABSs on the phase $\chi = \varphi_L - \varphi_R$. Evolution of the phase in time leads to Landau-Zener transitions, and thus fluctuations in the Josephson current. (f) Schematic illustration of a three-terminal Josephson junction with a wide central contact. Since the contact is much wider than the coherence length Cooper pairs cannot be formed by electrons from opposite sides (crossed Andreev reflection is suppressed) and thus quartets cannot be formed. Only single pair AC Josephson current can flow between $S_M$ and $S_L$, $S_R$.

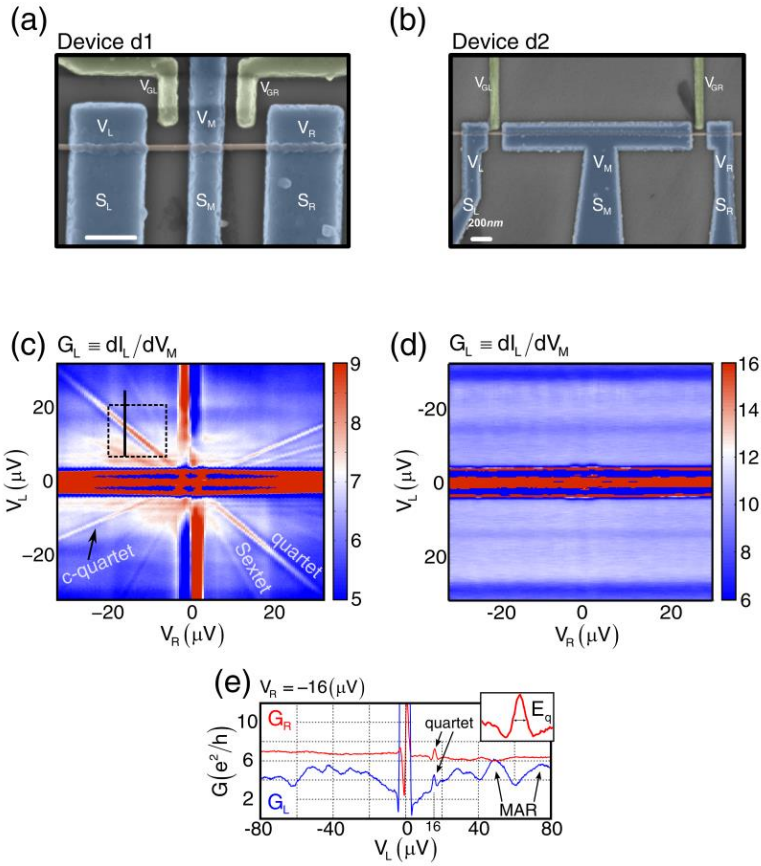

Fig. 2

**Fig. 2. Devices and differential conductance results**

(a) SEM image of device *d1*, scale-bar 300*nm*. The central superconducting contact is 200*nm*, namely, on the same order of magnitude as the coherence length. The gates (in green) were used to tune the transmission of the junction. (b) SEM image of device *d2*, scale-bar is of length 300*nm*. The central superconducting contact is 3*μm* wide, much larger than the coherence length (c) $G_L$ as a function of $V_L$ and $V_R$ measured in device *d1*. The quartet line, as well as other expected diagonal lines, is clearly seen. The solid line and dashed square are guidelines to Fig. 3A upper panel and Fig3B. (d) $G_L$ as function of $V_L$ and $V_R$ measured in device *d2*. No diagonal lines are observed. (e) $G_L$ (blue) and $G_R$ (red) as a function of $V_L$ in *d1*. The shape of the quartet peak, which resembles the Josephson current with the two side dips, is shown in the upper right corner with the quartet energy, Eq, indicated.

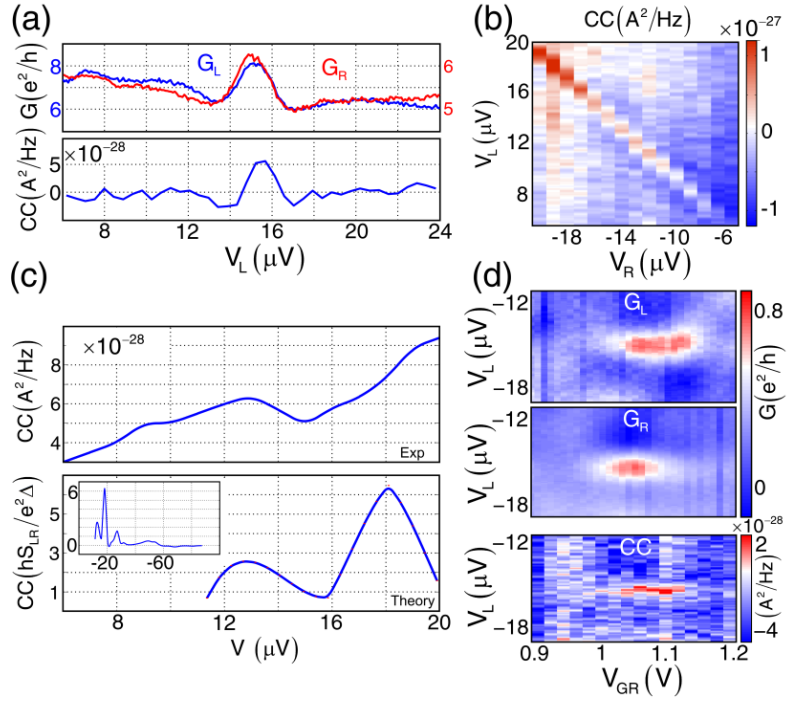

**Fig. 3**

**Fig. 3. Cross-correlation of current fluctuations and nonlocal conductance measurements.**

(a) Upper panel: Differential conductance cuts of $G_L$ and $G_R$ along the solid line in Fig 2C. Lower panel: Cross-correlation (CC) of current fluctuations at the left and right terminals. (b) CC as a function of $V_L$ and $V_R$ in the region defined by the dashed square of Fig 2C. (c) Upper panel: CC along the quartet line. Lower panel: Theoretical calculation of the CC. The maxima are due to Landau-Zener resonances. Inset: zoom-out in the bias voltage range. It should be noted that the measured CC in the experiments also drops after 20μV. (d) Upper and center panel: $G_L$ and $G_R$, respectively, as a function of the left contact bias, $V_L$, and the right gate voltage, $V_{GR}$. Lower panel: The CC as a function of $V_{GR}$.

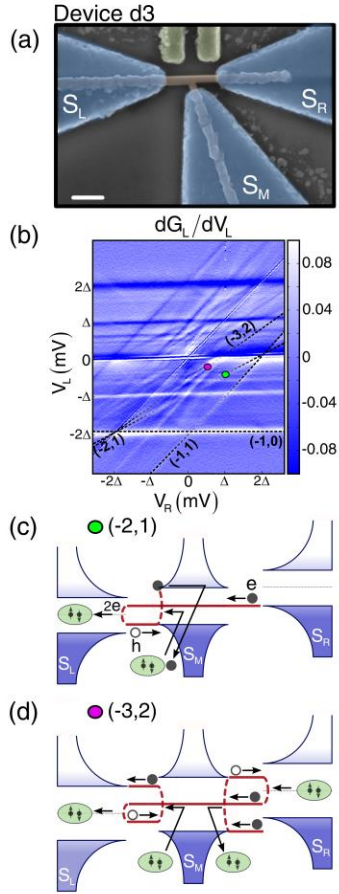

Fig. 4

**Figure 4. Local and nonlocal Multiple Andreev Reflections (MAR) in a three terminal Y-shape Josephson junction.** (a) SEM image of device d2, scale-bar is *200nm*. (b) $G_L$ as a function of $V_L$ and $V_R$. The lines labeled (-1,0) and (-1,1) correspond to first order *local* MAR. The lines labeled (-2,1) and (-3,2) are second and third order *nonlocal* MAR processes. (c) Schematic illustration of the (-2,1) process – second order (single Andreev reflection). (d) Schematic illustration of the (-3,2) process – third order (two Andreev reflections).

# Nonlocal supercurrent of quartets in a three-terminal Josephson junction

**Authors:**

*Yonatan Cohen[†,1], Yuval Ronen[†,1], Jung-Hyun Kang[1], Moty Heiblum[#,1], Denis Feinberg[2,3], Régis Mélin[2,3], and Hadas Shtrikman[1]*

**Affiliation:**

[1] *Braun Center for Submicron Research, Department of Condensed Matter Physics, Weizmann Institute of Science, Rehovot 76100, Israel*

[2] *CNRS, Institut NEEL, F-38042 Grenoble, France*

[3] *Université Grenoble-Alpes, Institut NEEL, F-38042 Grenoble, France*

[†] *Equal contributions*

[#] *Corresponding Author (moty.heiblum@weizmann.ac.il)*

**Methods and Supplementary Information:**

In this Supplementary Section we add details to the main text. We include a brief review of the theoretical background as well as the simulation method and results, as well as more information on the conductance and noise measurements.

## S1 - Theoretical Model

*A. General*

Let us consider a normal region connected to three superconducting terminals. When terminals $S_{L,R}$ are biased respectively at voltages $V_{L,R}$ with respect to terminal $S_M$, a coherent stationary motion of Cooper pairs occurs when $nV_L+mV_R=0$, where $(n,m)$ are integers. This involves $n$ pairs crossing from $S_M$ to $S_L$ and $m$ pairs crossing from $S_M$ to $S_R$ in a single quantum process [1, 2]. This multi-pair process unveils a phase combination $\varphi_{n,m} = n\varphi_L + m\varphi_R - (n+m)\varphi_M$ which, owing to the Josephson relation, $\frac{d\varphi_i}{dt} = \frac{2eV_i}{\hbar}$; ($i=L,R,M$), is a constant of motion. The main anomaly reported in the experiment along the line $V_L+V_R=0$ corresponds to a quartet (a pair of pairs) crossing from $S_M$ towards $S_{L,R}$, revealing the stationary phase $\varphi_q = \varphi_{1,1} = \varphi_L + \varphi_R - 2\varphi_M$. Sextet lines are also visible, though fainter, where, $(n,m) = (1,2)$ or $(2,1)$. These DC modes manifest static phase coherence despite the non-equilibrium conditions. Due to energy conservation, multi-pair processes are non-dissipative, contrary to the usual quasiparticle multiple Andreev reflections (MAR). Along the line $V_L=-V_R=V$, theory predicts that the quartet current $I_q(\varphi_q,V)$ is odd in phase and even in voltage. $I_q$ is similar to a DC Josephson supercurrent but it depends on $V$ as a new control parameter. It involves equal and perfectly correlated currents flowing through $S_L$ and $S_R$.

Choosing $\varphi_q = \varphi_L + \varphi_R - 2\varphi_M$ and $\chi = \varphi_L - \varphi_R$, as canonical variables one may as a first step begin with the Andreev bound state (ABS) energies at equilibrium $E_{ABS}(\varphi_q,\chi)$ which can be computed in a suitable model. Subsequently, one can use a semiclassical approximation and average out the drifting phase $\chi(t)$. This can be formally done by expanding $E_{ABS}(\varphi_q,\chi)$ in Fourier series in both variables keeping only the zeroth order component in $\chi(t)$. This leads to an effective energy $E_{eff}(\varphi_q)$, which is a function of $\varphi_q$ only. Then the average quartet current is found to be $I_{quartet}^{SC} = -\frac{2e}{\hbar}\frac{dE_{eff}}{d\varphi_q}$. This rough procedure reduces a set of two-dimensional ABS, valid at equilibrium, to a set of one-dimensional effective ABS. Yet, it neglects the quantum nature of the non-equilibrium processes, which take place as multiple Andreev reflections at the junction interface of all three superconductors. In the limit where the Josephson junction frequency $\omega_0 = \frac{2eV}{\hbar}$ is much smaller than the separation between the effective ABS, one obtains Landau-Zener transitions between the latter. None equilibrium Green's function calculations confirm this picture (see below and Figure S1a) and demonstrate that those transitions indeed induce a strong quartet noise.

## B. Results from non-equilibrium Green's function theory

The picture above is semi-phenomenological and a full non-equilibrium theory of transport is necessary. Such a theory is indeed available along the line $V_L=-V_R=V$; it involves the calculation of the Keldysh Green's function matrix $G(E,n)$, where $E$ is the energy and $n$ the index of the harmonics of the Josephson frequency $\omega_0 = \frac{2eV}{\hbar}$ [2, 3]. Voltages down to $0.1\Delta$ can be reached with about 100 harmonics. Mapping the full $(V_L,V_R)$ plane is out of reach, as independent Josephson frequencies $\omega_L$, $\omega_R$ would require much too large matrices. Results concerning a single dot model are found in Ref. [3]. The model used to describe the present experiment, instead, involves two single-level quantum dots $D_L$ and $D_R$ with energy levels $\varepsilon_L$ and $\varepsilon_R$ coupled to the terminals by couplings (broadening in the normal state) $\Gamma_L$, $\Gamma_M$ (for dot $D_L$), and $\Gamma_R$, $\Gamma_M$ (for dot $D_R$). For the purpose of interpreting the experiment, $\varepsilon_L$ and $\varepsilon_R$ are taken to be zero (resonant dots). Interactions are neglected owing to the large transparency. Figure S1a shows the quartet current flowing in terminal M and the cross-correlation noise $S_{LR}$. The $\Gamma$'s are taken as $\Gamma_L = \Gamma_R = 1.5\Delta$ and a smaller $\Gamma_M = 0.3\Delta$, owing to the finite width of the central superconducting finger that limits the crossed Andreev reflection.

Panels S1a, b show the quartet current and the crossed noise as a function of the quartet phase, fixing $eV = 0.15\Delta$ and taking into account a very small inelastic broadening $\eta = 10^{-6}$ [in units of $\Delta$] in the superconductors. A very strong resonance appears as marked dips at specific values of $\varphi_q$, that can be interpreted as resonant Landau-Zener transitions between two symmetrical ABS formed at zero voltage, triggered by the Josephson frequency $\omega_0$. This indeed resembles the effect of microwave irradiation on a quantum point contact [4]. Spectacularly, the cross correlation noise exhibits sharp peaks at the same phase values as the current dips (Fig S1a, Panel B). These peaks can be very high, signaling "trains" of quartets, in a way similar to the thermal noise due to transitions between a single junction ABS [5, 6]. Fig S1a, Panel C & D shows a broadening and an amplitude decrease in the current and the noise anomalies when increasing the inelastic parameter, where $\eta = 10^{-3}$. Panel S1a, E shows the variation with $V$ of the value of the cross correlation noise, calculated along the line $(V,-V)$ of the $G_L(V_L, V_R)$ map by taking into account thermal fluctuations (see Section C). First, one finds that the noise is positive. Second, its behavior is not monotonous, the first maximum being indeed due to the above Landau-Zener resonance. The maximum noise is much larger than $\frac{e^2\Delta}{h}$, indicating large bursts of quartets emitted within the Landau-Zener resonances. Those trends are also found in the experiment, where a non-monotonous variation of the maximum noise is obtained as well (Figure 3c, main text). No quantitative fit is attempted here, because the details of the current and noise variations with phase and voltage are very sensitive to the location of the resonances. In particular, i) the non-monotonous variation with $V$, with huge oscillations, and ii) the an-harmonic phase variation, with dips reflecting Landau-Zener transitions, are characteristic of such resonances and point towards the phase coherence of the quartet dynamics. Here, the parameters of the model are chosen to illustrate the main trends in a somewhat dramatic case. We also emphasize the extreme sensitivity of the quartet noise to the inelastic time, a parameter unknown in the experiment. As a last remark, measuring the charge 4*e* of quartets would require low transparency, making the detection much more difficult.

*C. Phase diffusion model close to the quartet line*

Here we present a semi-phenomenological picture which is capable of describing transport in the vicinity of the quartet line (*V*,-*V*), where no full microscopic solution is available anymore. In a voltage-biased junction, the Josephson supercurrent is probed indirectly through the shape of the conductance anomaly manifesting a rounded Josephson plateau in the *V*(*I*) characteristics. Its double-well shape can be described by an overdamped RSJ model [7]. The same is true here for the conductance anomaly, as a function of two voltages $V_L$, $V_R$. Transport by a quartet supercurrent is witnessed by a rounded plateau, centered on the quartet line. One can proceed and adiabatically describe the dynamics close to this line in the same spirit as the overdamped Josephson junction close to *V*=0, by means of an effective « quartet » RSJ model. This involves two branches in parallel: a quartet branch, non-dissipative and dependent on the phase $\varphi_q$, and a resistive branch. Setting, $V_L + V_R = v \ll |V_L|$, $|V_R| \approx V$, the phase $\varphi_q = \varphi_{q0} + \frac{2e}{\hbar} vt$ is a slow variable, while $\varphi_L - \varphi_R = \frac{4e}{\hbar} Vt$ is a fast one. The phase $\varphi_q$ evolves in an effective potential, which is determined here from the non-equilibrium Green's function calculation, by integrating the calculated quartet current $I_q(f_q) = \frac{2e}{\hbar} \frac{dU_{eff}}{d\varphi_q}$. Notice that this self-consistent procedure goes beyond the time-averaging procedure explained in section B1: One uses the microscopically exact solution on the quartet line to extrapolate to the slow adiabatic motion in its vicinity.

For this purpose, one can apply the theory of phase diffusion in the « washboard » potential formed by the quartet phase potential $U_{eff}(\varphi_Q) - \frac{\hbar I_q}{2e} \varphi_Q$, where $I_q$ is the average quartet current. Application of the Ambegaokar-Halperin phase diffusion model [8] yields the phase thermal probability distribution $p(\varphi_q)$ as well as the quartet current-voltage $I_q(v)$ characteristics. The conductance calculated from this scheme has the classical shape found in the usual Josephson effect and also in the present experiment on the quartet line (Fig S1b). It only depends on a single parameter $\gamma = \frac{\hbar}{2e} \frac{I_{Qc}}{k_B T}$, that can here be estimated from the universal shape of the anomaly to be about 1-2 at 30mK.

This argument confirms that the conductance anomaly across the quartet line underlies the quartet phase, and allows to evaluate a typical quartet energy to be about 60-100mK. This model also allows calculating the thermally averaged value of the crossed noise at the center of the anomaly ($v=0$). It is plotted in Figure S1a Panel E and can be much larger than $\frac{e^2 \Delta}{h}$. Yet, this model does not allow to fully calculating the crossed correlation noise anomaly across the quartet (V, -V) line, owing to the strong non adiabatic character of the quartet noise which dramatically depends on Landau-Zener transitions. Those transitions are not correctly described by the model described above.

*D. Nonlocal multiple Andreev reflections vs quartets.*

We now discuss the zero-energy nonlocal MAR process which might compete with the quartet mechanism along the line (V, -V). In the main text, we explain several observations that distinguish between the two effects. Perhaps the most important one is the sign of the crossed noise measured by correlating the current fluctuations on the left and right terminals. While the measured signal is positive, the crossed noise expected from the zero-energy nonlocal MAR process is negative. This can be first understood by an intuitive argument: in a zero-energy MAR –particles are transported between terminals at different voltages with the help of energy-conserving Cooper pair transitions. Such a fermionic dissipative transport is expected to result in anti-bunching, e.g. negative noise correlations between terminals at different voltages, thus negative CC between L, R terminals. This is indeed confirmed by a full Keldysh calculation, made with a single level dot model, taking into account all conserving MAR processes together with quartets [3]. Figure 5 of Ref. 3 shows an essential result: it compares the quartet current Ic in terminal C (the setup is symmetrical), the quasiparticle current (Ia-Ib) and the noise correlation CC, for different values of the voltage *V*. For large voltages the MAR current dominates the quartet current and the CC is indeed weak and negative (red and green curves in panel a, b, c of Figure 5, obtained for a resonant junction, as in the experiment). On the contrary, for small voltages, the MAR current is smaller than the quartet current and the CC is strong and positive (black and light blue curves). This phenomenon is generic and observed for a more realistic two-level dot model as well.

**S2 - Measurement Setup**

The experimental setup is shown in Fig. S2a. Resonance frequencies of the two LC circuits were matched in order to enable the cross-correlation measurements at ~705KHz.

*A. Differential conductance measurements:*

As described in the main text, differential conductance was measured by applying an input ac signal of 0.8µVrms at 705 KHz to the center contact, $S_M$, while measuring the differential voltages, $V_L$ and $V_R$, on the left and right contacts, $S_L$ and $S_R$, respectively. The 500Ω load resistors were chosen to be significantly lower than the typical values of the sample resistance so that they serve as effective drains pulling most of the current to the ground. We then define: $G_L=dI_L/dV_M$, $G_R=dI_R/dV_M$, where $I_L=V_L/500Ω$ and $I_R=V_R/500Ω$. Figure S2b presents a color plot of $G_L$ and $G_R$ as a function of the applied biases $V_L$ and $V_R$ in *device d1*.

*B. Cross-correlation of current fluctuations measurements:*

In the cross-correlation of current fluctuations measurement no AC signal is applied. DC bias voltages, however, produce current fluctuations, $dI_L$ and $dI_R$ (ac component at relatively low frequencies ~ 705kHz). We are interested in the cross correlation of the current fluctuations $<dI_LdI_R>$. The current fluctuations introduce voltage fluctuations $dV_L=dI_L•500Ω$ and $dV_R=dI_R•500Ω$ at the inputs of a home-made, cold (1K) amplifier (the gains of which were measured in advance to be $g_L$=6.12 and $g_R$=5.77). Another amplification stage was used at the output of the dilution fridge using NF amplifiers each with a gain of 200. Both signals are multiplied and amplified by a home-made cross correlator with a central frequency of 730KHz, resolution band width of RBW=100KHz and gain of $g_{CC}^2 = 10^7$. Finally, the cross correlator signal undergoes an RC filter. The CC can be estimated by:

$$CC_{tot} = \langle (dI_L \times 500 \times g_L \times g_{NF} \times g_{CC})(dI_R \times 500 \times g_R \times g_{NF} \times g_{CC}) \rangle \times RBW$$
$$= \langle dI_L dI_R \rangle \times [500^2 \times g_L \times g_R \times g_{NF}^2 \times g_{CC}^2 \times RBW]$$
$$= \langle dI_L dI_R \rangle \times \alpha$$

However, parasitic effects such as RF picked up by both output lines, cross talk coming from capacitance between the output lines etc., add an independent "background" cross correlation, $CC_{tot} = \langle dI_L dI_R \rangle \times \alpha + CC_0$.

Since the load resistor was chosen to be very small (500Ω) relative to the sample resistance, the voltage signal is very small, relative to the background cross correlation. Hence, the background must be calibrated and subtracted as explained in the next section.

## S3 − Cross-correlation calibration

To calibrate the background, before each measurement of $V_L$ where we scan the cross correlation (as we move through the quartet line), we perform the same measurement at a high magnetic field of $B$=200mT (above the critical field of the SCs so that all contacts are in the normal state). At zero bias voltages, no current flows through the device and we expect the voltage fluctuations $dV_L$ and $dV_R$ to be uncorrelated. Hence, we take the cross correlation measured at this high magnetic field and at zero bias as our background cross correlation. An example of such cross correlation measurement is shown in Fig. S3.

## S4 – Negative cross-correlation on the complementary quartet lines

As mentioned in the main text we expect to observe a positive cross-correlation of the current fluctuations, between the left and right terminals, along the quartet conductance line. As a sanity check we measured the CC along different processes where we expect to get negative cross correlation.

In Fig. S4A we illustrated a quartet process which is named complementary quartet process, which is merely a permutation of the terminals from the process described in the main text. In this process Cooper pairs from the left and center contact enter the right contact and in the process they are entangled between themselves. This process is thus called a complementary quartet process. In Fig S4B we sketched the complementary quartet ABS which is the mechanism for the creation of a quartet in the right contact.

The results of this process are shown in Fig S4C. We placed the right terminal at a bias of $V_R = -13\,(\mu V)$ and measured the differential conductance in the left and right contact (upper and middle panels) and the cross correlation of the current fluctuations in the left and right terminals. Concentrating on the blue shaded region (where the complementary quartet process occurs) a clear reduction in the cross correlation is observed – originating from a negative contribution of the process. Concentrating now on the red shaded area, which is the region where a trivial supercurrent flows from the left to the right terminal, once again a clear reduction of the cross correlation is observed.

## S5 – Confirming the absence of a common element in *device d2* and *device d3*

To demonstrate experimentally that there is no normal element which is common to the two junctions in *device d2* and *device d3*, we performed CC measurements in the normal state of the devices, at high magnetic field. A fixed voltage $V_R$ = -14 µV was applied to the left contact, $S_L$, and the CC of the current fluctuations on $S_L$ and $S_R$ was measured as a function of $V_L$. As shown in Fig. S5A, in the presence of a common element, the current $I_R$ going to $S_R$, would be correlated to the current $I_L$ coming from $S_L$. Thus, a finite CC would be measured. In contrast, in the absence of such common element (or if this element is significantly smaller than the other two, $R_C \ll R_L, R_R$), the currents $I_L$ and $I_R$ are completely independent and the CC would vanish, as seen in Fig. S5B. The same argument evidently holds in the tunneling regime as well, where instead of three resistive elements, there are three tunnel junctions. If the probability, $\Gamma_C$, of tunneling from the center of the junction to the central contact $S_M$, is comparable to the other two tunneling probabilities, then the currents $I_L$ and $I_R$, would be correlated. However if $\Gamma_C \ll \Gamma_L, \Gamma_R$, then the CC would vanish.

The measurement result is shown in Fig. S5E. Indeed we measured zero CC which demonstrates the absence of the common element. Even if this leaves room for a very small common element which gives rise to CC smaller than our measurement error, this common element is clearly much smaller than it is in the case of the T-shaped *device d3*. However, the conductance anomaly on the $V_L=-V_R$ line in *device d1* is very similar to the one in *device d3*. Therefore, we conclude that such common element cannot be the origin of the conductance anomaly.

## S6 – MAR and quartet in *device d3* (T-shaped nanowire device)

Fig. S6 shows $G_L$, of *device d3* as a function of $V_L$ and $V_R$. To better identify the MAR lines, which are faint relative to the supercurrent lines, the derivative of the differential conductance, $dG_L/dV_L$, was taken and is shown in Fig. S6B. Both MAR lines as well as supercurrent lines can be seen. The figures emphasize the similarity of the quartet line ($V_L=-V_R$) to the other supercurrent lines ($V_L=0$, $V_L=V_R$) and the difference from MAR lines (*e.g.* $V_L=-2\Delta$).

**Figures**

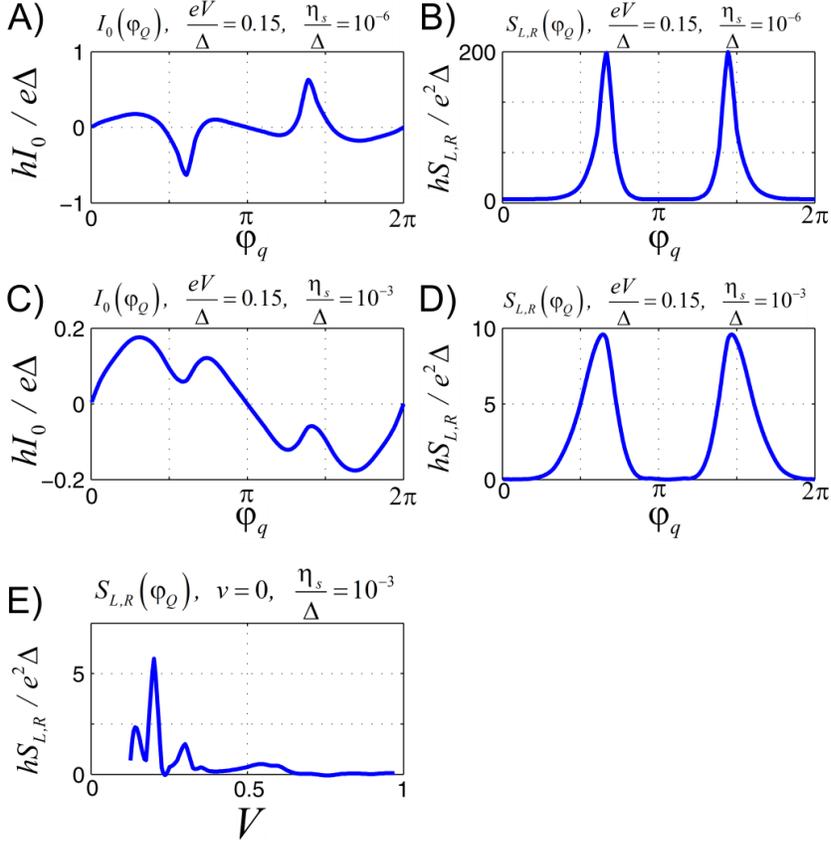

**Fig. S1a. Two-dot model.**

NEFG calculation of the quartet current $I_0$ (in the central terminal M) and of the current cross correlation (CC) $S_{LR}$, as a function of the quartet phase $\varphi_q$ (panels A-D) and of the voltage (panel E), with $eV = 0.15\Delta$, the inelastic parameter $\eta_s = 10^{-6}\Delta$ (panels A, B) and $\eta_s = 10^{-3}\Delta$ (panels C, D, E). Panel A shows two resonances corresponding to Landau Zener transitions between the adiabatic quartet states. At the same phase values, the CC displays sharp maxima with very high values. Both trends are still present with a larger inelastic parameter (panels C, D) but broader and reduced amplitude. Panel E shows the non-monotonous variation of the CC at the center of the quartet line but sweeping the voltage V, showing that the resonances occur at certain V values. Thermal fluctuations are taken into account with gamma = 0.5 (see text).

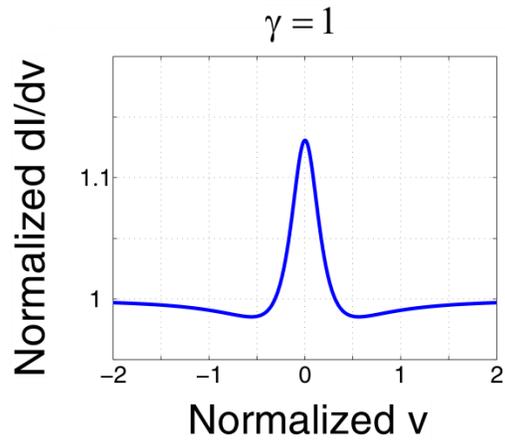

**Fig. S1b.** Differential conductance of the quartet

Differential conductance as a function of voltage v, taken from the center of the quartet line, in units of the maximum quartet current, the normal state conductance being taken to one. It is calculated from the effective phase diffusion model, plugging in it the results of the Green's function calculation for IQ($\varphi_Q$)

a)

b)

c)

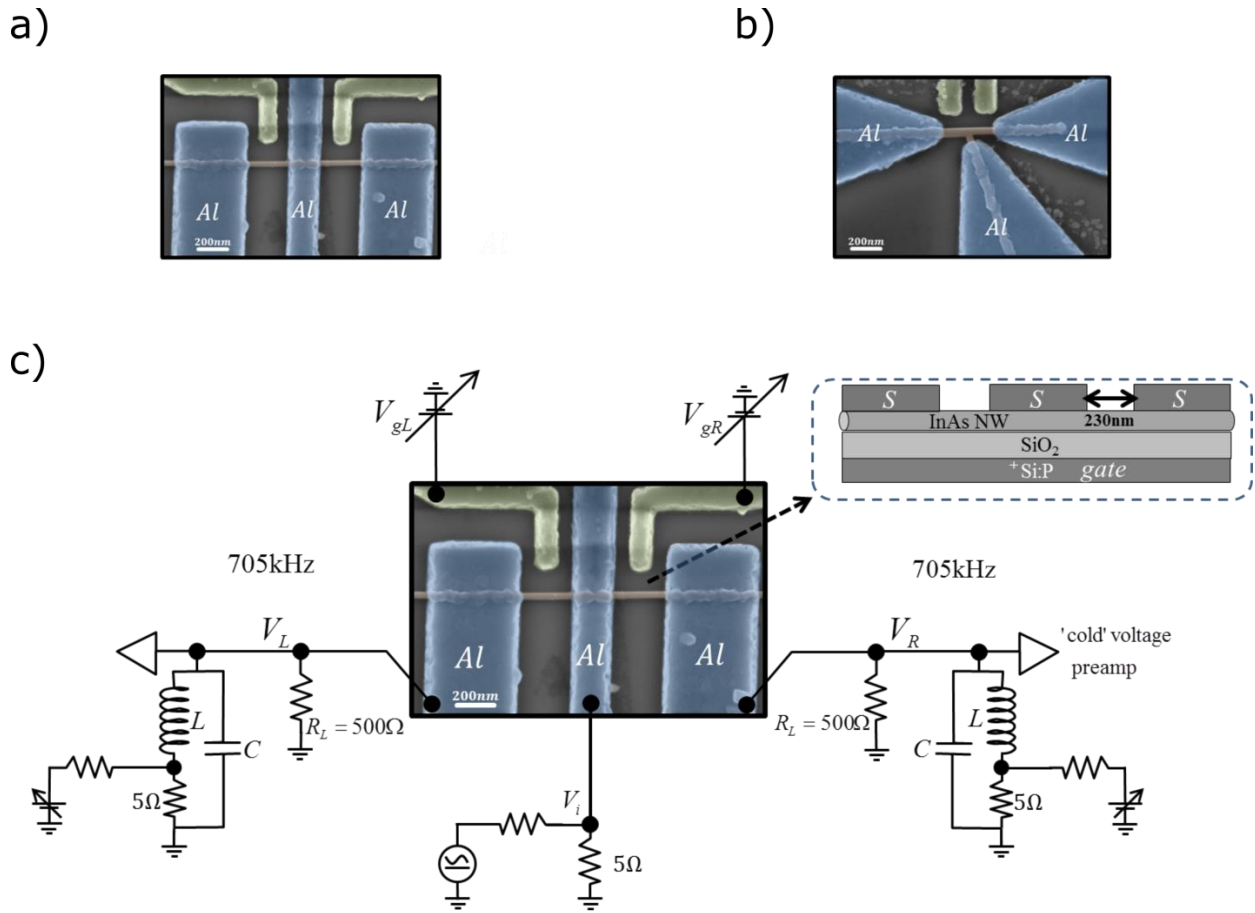

**Fig. S2a. Devices and Measurement Setup.**

A) D1 configuration. Three superconducting contacts are placed on a single nanowire. The central contact is made much narrower than the coherence length of the superconductor so that crossed Andreev reflection is allowed. B) *Device d3*. Three superconducting contacts are placed on the three branches of a T-shaped nanowire. Compared to the configuration of *device d1*, here there is much higher probability for direct transport of quasiparticles and Cooper pairs from the left contact to the right contact. C) Our measurement setup. The differential conductance and the cross correlation of current fluctuations were measured using this setup as described above. The device configuration is schematically illustrated at the top right hand corner.

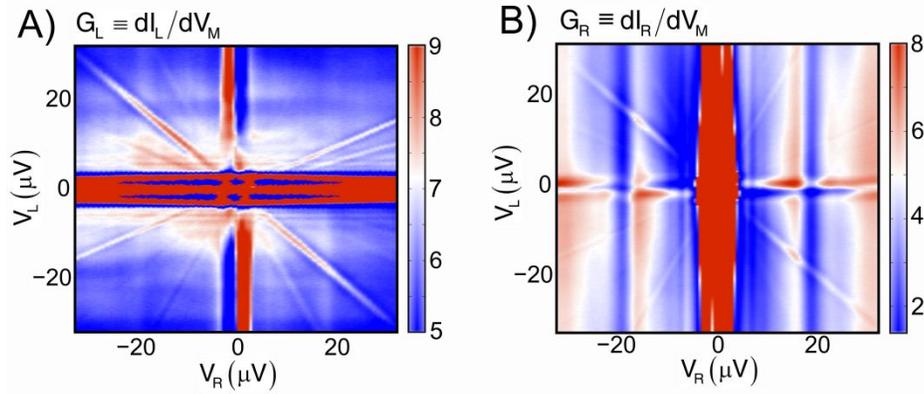

**Fig. S2b. Differential conductance measurements results**

A) $G_L$ vs. $V_R$ and $V_L$ measured in *device d1*. B) $G_L$ vs. $V_R$ and $V_L$ measured in *device d1*.

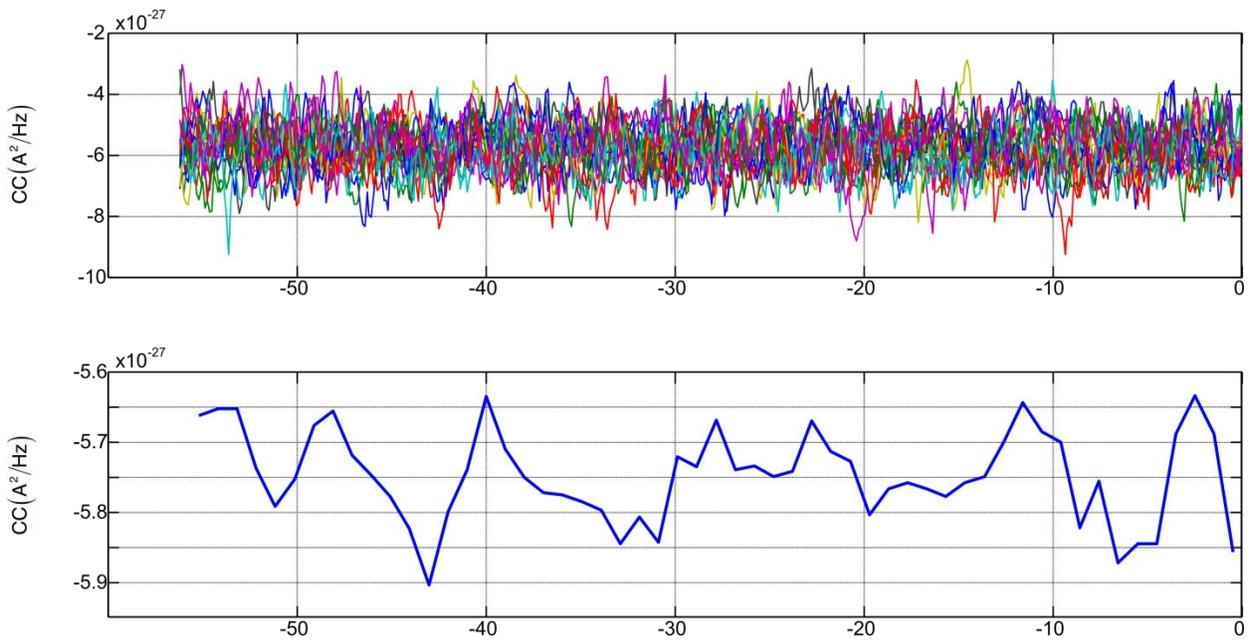

**Fig. S3. Cross correlation calibration.**

Cross correlation as a function of $V_L$, measured at B=200mT, well above the critical magnetic field of aluminum. At zero bias, there is no current and therefore we expect zero cross correlation. It can be seen that even at a finite bias of 50µV the cross correlation is essentially the same as at zero bias (within an uncertainty that is significantly smaller than the cross correlation measured at B=0 on the quartet resonance). This is due to the fact that in *device d1*, the central contact disconnects the two sides of the device, or in other words most of the current from the left/right contact flows to the central contact rather than to the other side.

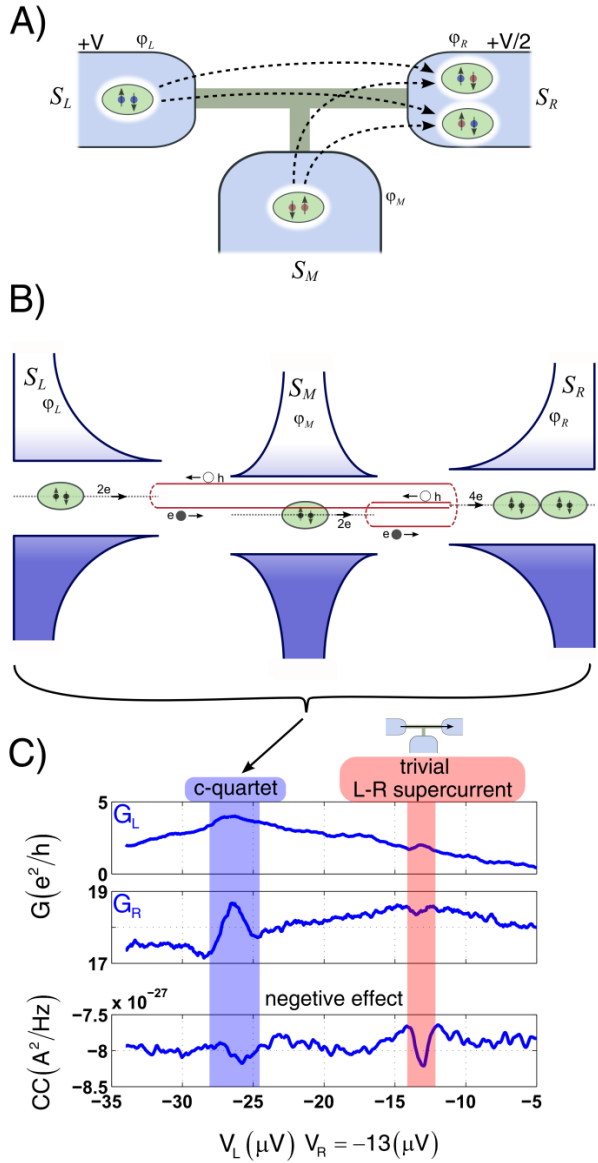

**Fig. S4. Complementary quartet lines.**

A) Schematic illustration of the 'right' (right contact) complementary quartet line. B) Schematic illustration of the 3-terminal complementary quartet ABS. C). The upper and middle panels show the differential conductance in the left and right terminals. The lower panel displays the cross correlation of the current fluctuation, showing a negative peak in the location of the complementary quartet.

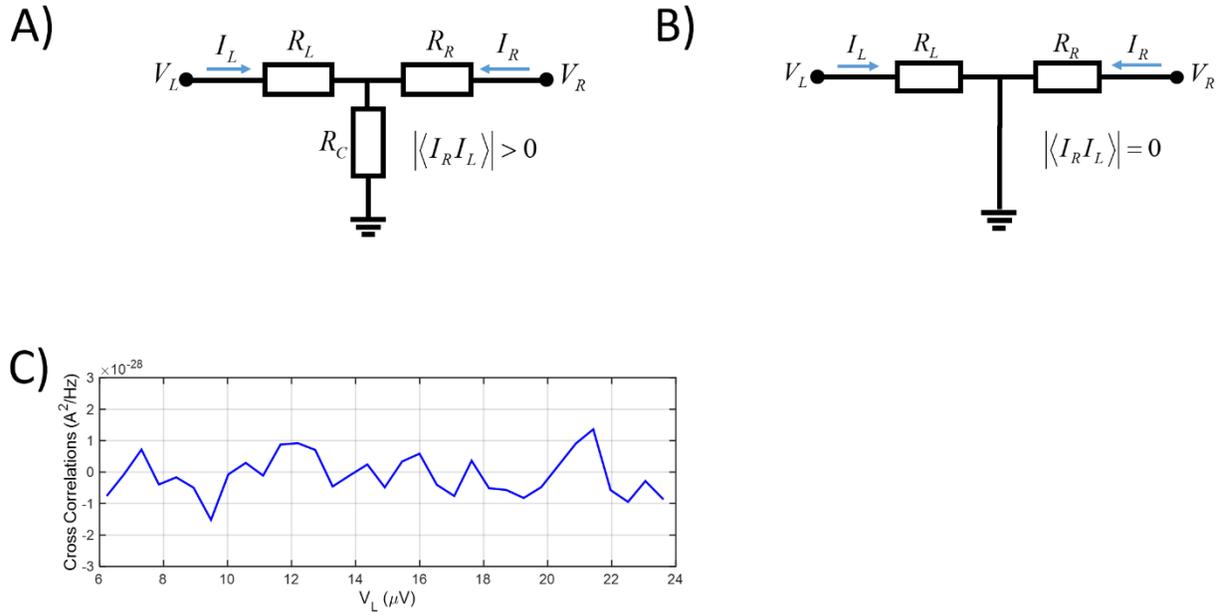

**Fig. S5. Absence of a common element.** A) Common element, Rc, is comparable in size to $R_L$ and $R_R$: cross-correlation of current fluctuation is expected to have a positive sign. B) No common element, Rc is comparable in size to $R_L$ and $R_R$: cross-correlation goes to zero. C) Zero CC as a function of $V_L$ proving scenario B is our case.

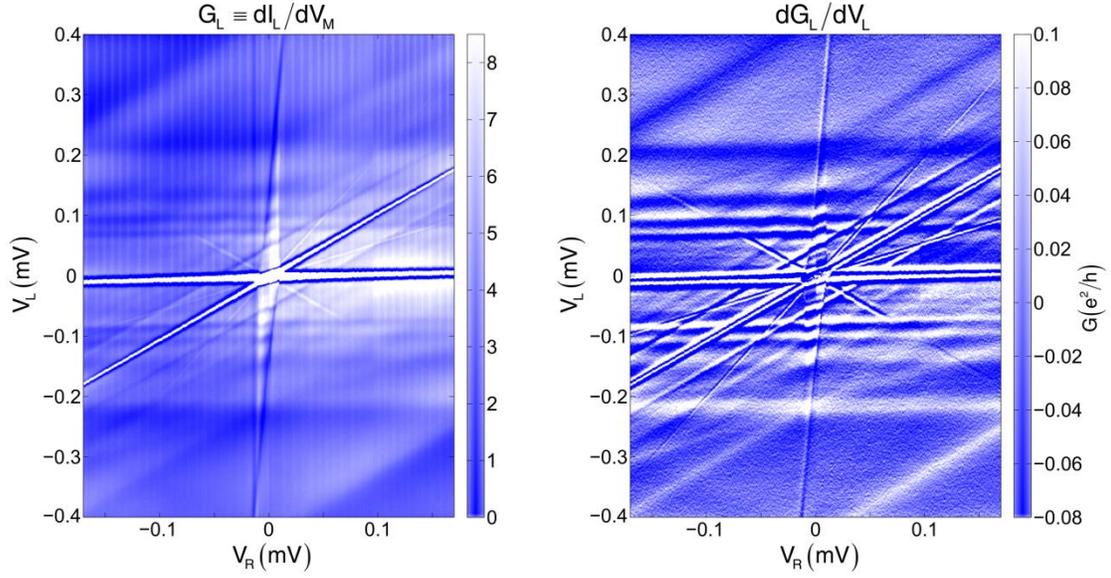

**Fig. S6. MAR and quartet.**

A) $G_L$, of *device d3* as a function of $V_L$ and $V_R$. B) To better identify the MAR lines, which are faint relative to the supercurrent lines, the derivative of the differential conductance, $dG_L/dV_L$ is plotted as a function of $V_L$ and $V_R$. Both the MAR lines as well as the supercurrent lines can be seen. The figures emphasize the similarity of the quartet line ($V_L=-V_R$) to the other supercurrent lines ($V_L=0$, $V_L=V_R$) and the clear difference with respect to MAR lines (*e.g.* $V_L=-2\Delta$).